\documentclass[11pt]{iopart}
\usepackage{iopams}
\usepackage{setstack}
\usepackage[dvips]{graphicx}
\usepackage{epsfig}
\begin{document}
\title{Electrical networks on $n$-simplex fractals}

\author{R. Burioni$^{1,2}$, D. Cassi$^{1,2}$, and F.M. Neri$^1$}
\address{$^1$ Dipartimento di Fisica, Universit\`a degli Studi di
Parma, viale Usberti 7/A, 43100 Parma, Italy}
\address{$^2$ INFN, Gruppo
Collegato di Parma, viale Usberti 7/A, 43100 Parma, Italy}
\address{E-mail: burioni@fis.unipr.it, cassi@fis.unipr.it, and neri@fis.unipr.it}

\begin{abstract}
The decimation map $\mathcal{D}$ for a network of admittances on an
$n$-simplex lattice fractal is studied. The asymptotic behaviour of
$\mathcal{D}$ for large-size fractals is examined. It is
found that in the vicinity of the isotropic point the eigenspaces of
the linearized map are always three for $n \geq 4$; they are given a
characterization in terms of graph theory. A new anisotropy
exponent, related to the third eigenspace, is found, with a value
crossing over from $\ln[(n+2)/3]/\ln 2$ to
$\ln[(n+2)^3/n(n+1)^2]/\ln 2$.
\end{abstract}

%
%
%
%
%
%
%
%
%
%
%
%
%
%
%
%
%
%
%
%
%
%
%
%
%
%
%
%
\section{Introduction}
\label{Sec:Introduction} The ``$n$-simplex lattice'', built through
the iteration of a complete graph $K_n$, is the generalization of
the usual two-dimensional Sierpi\'{n}ski gasket fractal to a
$d$-dimensional space, with $n=d+1$. It was first introduced thirty
years ago by Dhar \cite{Dhar1977} (in a slightly different,
``truncated'' version) as one of the first examples of lattices with
nonintegral dimensionality.\\
The problem of \textit{electrical networks} on $n$-simplex lattices
was initially considered under the simple form of \textit{isotropic
resistor networks} \cite{Gefen1981}, modelling the conducting
backbone of a percolating system at criticality. The aim was to
study the anomalous scaling law for the total conductance $\sigma$
(or complex admittance, in the case of a general passive element) as
a function of the size: it was found that $\sigma\sim
L^{-\bar{\zeta}}$, with $\bar{\zeta}=\ln[(n+2)/n]/\ln 2$. It was
soon realized \cite{Vannimenus1984} that on these fractals the
macroscopic conductance is isotropic even starting from microscopic
anisotropy. Hence, \textit{anisotropic} networks were considered, in
order to find the scaling exponent $\bar{\lambda}$ with which the
conductance anisotropy (defined as the difference from 1 of the
ratio of two conductances $\sigma_\alpha$ and $\sigma_\beta$
measured along different directions) vanishes:
$(\sigma_\alpha-\sigma_\beta)/\sigma_\beta\sim
L^{-\bar{\lambda}}$. \\
The problem consists in finding the decimation transformation
$\mathcal{D}$ connecting the conductances of generation $g+1$ of a
fractal to those of generation $g$, then linearizing $\mathcal{D}$
near the isotropic point. The largest eigenvalue of the linearized
map, $e_1=n/(n+2)$, is then related to the isotropic exponent by
$\bar{\zeta}=-\ln e_1/\ln 2$; the second eigenvalue $e_2$ is related
to the anisotropy exponent by $\bar{\lambda}=\ln(e_1/e_2)/\ln 2$.
Since no simple method is known for finding $\mathcal{D}$ for $n>3$,
the analytical challenges in the anisotropic case are much greater.
The greatest effort in this direction at that time was that by
Vannimenus and Kne\v{z}evi\'{c} \cite{Vannimenus1984}, who managed
to find $\bar\lambda$ for $n=4$, $n=5$, and suggested the general
form $\bar{\lambda}=\ln[(n+2)/(n+1)]/\ln 2$ (corresponding to
$e_2=n(n+1)/(n+2)^2$). More recently, in the light of a renewed
interest in the restoration of isotropy in several models defined on
fractals \cite{Barlow1995}, Jafarizadeh \cite{Jafarizadeh2000}
succeeded in proving the formula suggested in
\cite{Vannimenus1984} for $n$-simplex lattices for any $n$.\\
The general problem of the decimation of a network of conductances
on the $n$-simplex lattice had never been completely solved up to
now. In a previous paper \cite{Burioni2005} we found the exact map
for the decimation of a network of impedances on a three-dimensional
Sierpi\'nski gasket (4-simplex lattice), with a method based on the
direct manipulation of the Laplacian matrix of the circuit. In this
paper we generalize that method and show a detailed procedure to
find the decimation map $\mathcal{D}$ for arbitrary $n$. Instead of
conductances, more general complex admittances will be used. For the
first time, a systematic analysis of the asymptotic expansion of
$\mathcal{D}$ near the isotropic fixed point will be given. \\
The main results are the following. There are always exactly
\textit{three} eigenvalues regardless of $n$, that is, of the
dimension of the fractal (being two just for the two-dimensional
system). The third eigenvalue is $e_3=3n/(n+2)^2$ and is related to
a secondary anisotropy exponent equal to $\ln[(n+2)/3]/\ln 2$; however,
this exponent crosses over to $\ln[(n+2)^3/n(n+1)^2]/\ln 2$ for
large systems. The eigenspaces corresponding to $e_2$ and $e_3$ will
be studied for the first time. We will show that the eigenvectors
corresponding to $e_2$ are those that allow a direct simplified
(mesh-to-star) treatment of the problem. It will also be found that
the eigenspace corresponding to $e_3$ is related to the space of
cycles of even length on the complete graph $K_n$, and has the
highest dimensionality for $n$ large, scaling as $n^2$.\\
The plan of the paper is as follows. In section
\ref{Sec:the_basic_cell} we introduce the model and the notation we
will use throughout the paper. In section \ref{Sec:Decimation} we
describe our general decimation procedure, which we apply first to
case of the isotropic fractal and, second, to that of an isotropic
fractal with a small perturbation, which provides us with the linear
expansion of the decimation map $\mathcal{D}$ in the vicinity of the
isotropic point. Section \ref{Sec:Eigenvalues} examines the
eigenvalues and eigenspaces of the linearized map. Section
\ref{Sec:Conclusions} contains our conclusions.
%
%
%
%
%
%
%
%
%
%
%
%
%
%
%
%
%
%
%
%
%
%
%
%
%
%
%
%
\section{Basic cell and notations. Construction of a network of admittances. Statement of the problem}
\label{Sec:the_basic_cell}

\begin{figure}[t]
\begin{center}
\includegraphics[width=0.7\textwidth]{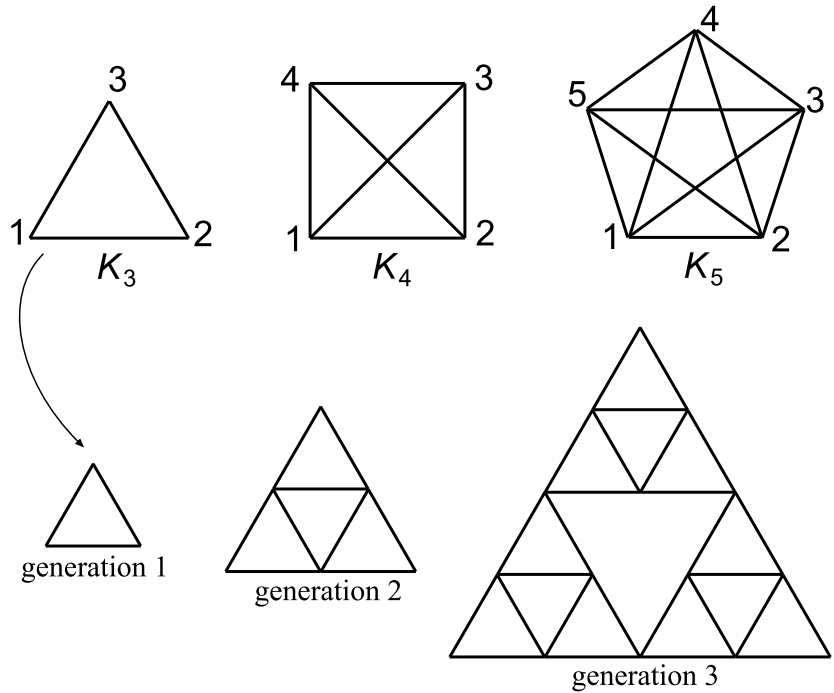}
\end{center}
\caption{{\it Top:} three $K_n$ complete graphs. The $n$-simplex lattice is a
fractal built using a $K_n$ as the basic cell. {\it Bottom:} the
construction of the two-dimensional Sierpi\'{n}ski gasket, or
$3$-simplex lattice, starting from the basic cell.}
\label{fig:Kn_graphs}
\end{figure}

\begin{figure}[t!]
\begin{center}
\includegraphics[width=0.7\textwidth]{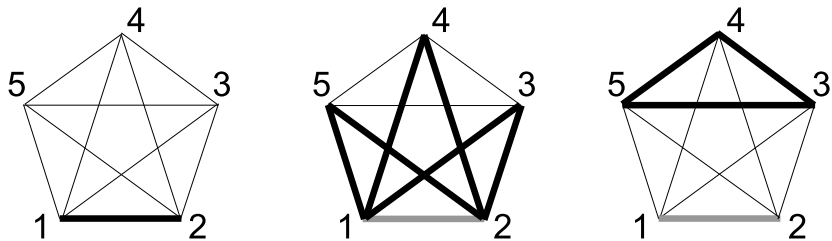}
\end{center}
\vspace{-1pt} \caption{\small We consider the $K_n$ graph and focus
on one of its links (for example link $(1,\,2)$ in graph $K_5$ on
the left). The number of adjacent links is $2(n-2)$ (middle): $n-2$
incident in vertex 1 and $n-2$ incident in vertex 2. The number of
non-adjacent links is $1/2\,(n-2)(n-3)$ (right), that is the number
of links of the $K_{n-2}$ subgraph obtained by removing vertices 1
and 2 from $K_n$.} \label{fig:adjacentlinks1}
\end{figure}

The $n$-simplex lattice is built by the iteration of a basic cell
(figure \ref{fig:Kn_graphs}). The basic cell is the complete graph
$K_n$ ($n$-simplex), with $n$ vertices and $n(n-1)/2$ links, one between each pair
of vertices. We follow the convention to arrange the vertices of the
graph as those of an $n$-polygon and label them with the numbers
from $1$ to $n$ counterclockwise.
We recall that two links are called \textit{adjacent} when they
share a vertex, and \textit{non-adjacent} when they don't. In a
$K_n$ graph (see figure \ref{fig:adjacentlinks1} for graph $K_5$)
each link has exactly $2(n-2)$ adjacent links and $1/2\,(n-2)(n-3)$
non-adjacent links (corresponding to a $K_{n-2}$ complete subgraph).\\
%
%
%
%
%
%
%
%
%
%
%
%
%
%
%
%
%
%
%
%
%
%
%
%
%
%
%
%
%
\subsection{Construction of a network of admittances. Kirchhoff's laws for the basic cell.}

We consider first a {\it general} electrical network, that can be
represented by a graph, i.e., a set of nodes connected by links,
with an electrical pole on each node $i$ and an electrical element
on each link. We will consider here only linear passive elements,
that is resistances, inductances and capacitances. We call
$z_{(i,\,j)}$ the impedance of the element (if any) connecting nodes
$i$ and $j$, and $\sigma_{(i,\,j)}=z_{(i,\,j)}^{-1}$ its admittance.
$V_i$ is the potential at node $i$ and $I_i$ the external current
incoming at node $i$. The key formula we will use in the following
is Kirchhoff's node law:
\begin{displaymath}
I_{i}=\sum_{k\neq i}\sigma_{(i,\,k)}(V_{k}-V_{i})=
\sum_{k\neq i}\sigma_{(i,\,k)}V_{k}-\left(\sum_{k\neq i}\sigma_{(i,\,k)}\right)\,V_{i},
\end{displaymath}
that can be expressed in matrix form:
\begin{equation}
\vec{I}=\mathbf{L}\,\vec{V}, \label{eq:external_currents_general}
\end{equation}
where $\vec{V}=(V_1,V_2,\ldots V_n)^\mathrm{T}$,
$\vec{I}=(I_1,I_2,\ldots I_n)^\mathrm{T}$, and $\mathbf{L}$  is the
Laplacian matrix of the circuit (see e.g. \cite{Harary1969}, and
\cite{Wu2004} for a recent application): $L_{i\,i}=\sum_{j\neq
i}\sigma_{(i,\,j)}$ for diagonal entries, and
$L_{i\,j}=-\sigma_{(i,\,j)}$ for nondiagonal entries. Since the
Kirchhoff's law for currents states that $\sum_{i}I_i=0$, the
entries of $\mathbf{L}$ are linearly dependent, and
$\det(\mathbf{L})=0$.\\
\begin{figure}[t!]
\begin{center}
\includegraphics[width=0.5\textwidth]{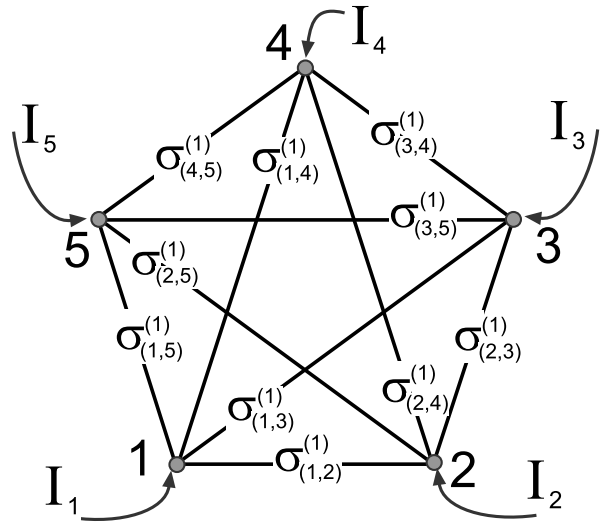}
\end{center}
\caption{The $K_5$ graph and the labelling used. External potentials are not shown.}
\label{fig:K5_basiccell}
\end{figure}
We now consider the $K_n$ graph, the basic cell (or
$1^{\mathrm{st}}$-generation fractal) of the $n$-simplex lattice
(figure \ref{fig:K5_basiccell}). The admittance between pole $i$ and
pole $j$ is $\sigma_{(i,\,j)}^{(1)}$. The relation between external
currents and potentials is $\vec{I}=\mathbf{L^{(1)}}\,\vec{V}$,
where $\mathbf{L^{(1)}}$ is the Laplacian matrix of the
$1^{\mathrm{st}}$-generation fractal:
\begin{equation}\left\{
\begin{array}{ll}
\displaystyle{L^{(1)}_{i\,i}=\sum_{j\neq i}\sigma_{(i,\,j)}^{(1)}} & \mathrm{diagonal\;entries}\\[2.5ex]
\displaystyle{L^{(1)}_{i\,j}=-\sigma_{(i,\,j)}^{(1)}} & i\neq j\\[2.5ex]
\end{array}\right.
\label{eq:laplacian_1stgen}
\end{equation}
If we start from the Laplacian matrix, the admittances of the
network are given by its off-diagonal terms:
$\sigma_{(i,\,j)}^{(1)}=-L^{(1)}_{i\,j}$.
This relation will play an important role in our following calculations.\\
In several expressions in the following we will sum over
\textit{links} rather than over \textit{vertices}: thus, we will
find it useful to label links with numbers from 1 to $n(n-1)/2$,
rather than with pairs. In that case we will use greek letters:
$(i,\,j)\equiv\alpha$, with $\alpha=1,2,\ldots n(n-1)/2$; hence,
instead of $\sigma_{(i,\,j)}^{(1)}$, the form
$\sigma_{\alpha}^{(1)}$ will also be used.
%
%
%
%
%
%
%
%
%
%
%
%
%
%
%
%
%
%
%
%
%
\subsection{Construction of a network of admittances: higher generations and decimation map}
\label{Sec:general_procedure}
\begin{figure}[t!]
\begin{center}
\includegraphics[width=\textwidth]{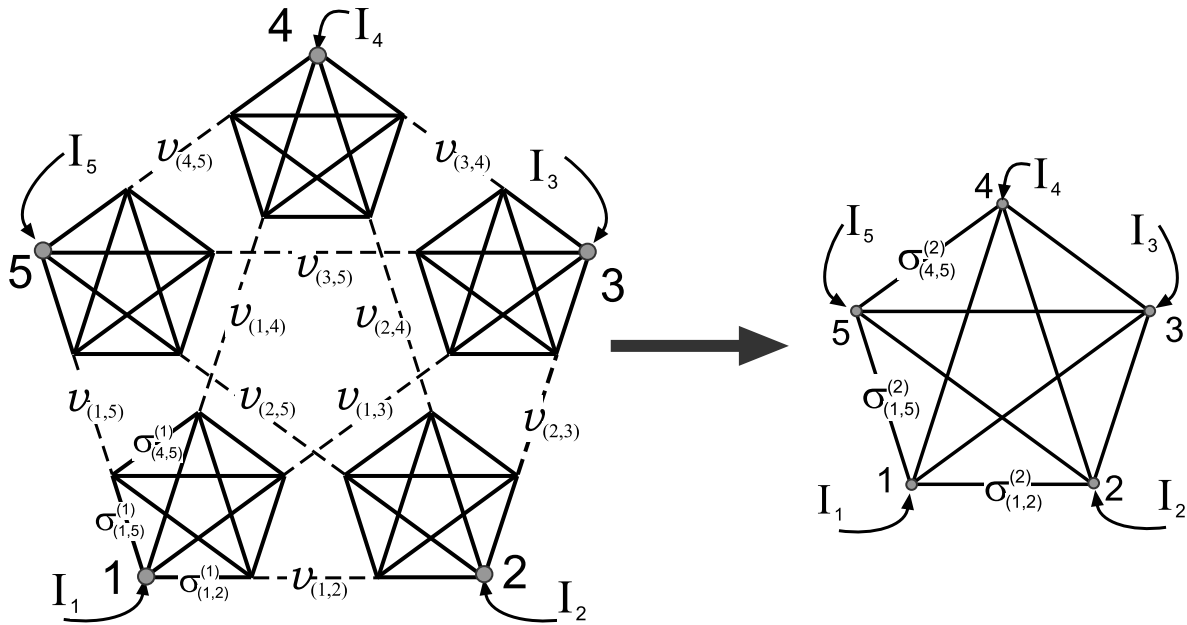}
\end{center}\vspace{-1pt}
\caption{{\it Left:} construction of the
$2^{\mathrm{nd}}$-generation fractal starting from a basic cell with
admittances $\sigma_{(i,\,j)}^{(1)}$. Dashed lines denote short
circuits between poles, coinciding with internal poles. The internal
potentials are called $v_{(i,\,j)}$; they are in a one-to-one
correspondence with the links of the basic cell. The external
potentials are not shown. {\it Right:} the circuit built on the
$2^{\mathrm{nd}}$-generation fractal is equivalent to that built of
a $1^{\mathrm{st}}$-generation fractal with new admittances
$\sigma_{(i,\,j)}^{(2)}$.} \label{fig:K5_generazione2}
\end{figure}
\begin{figure}[t]
\begin{center}
\includegraphics[width=\textwidth]{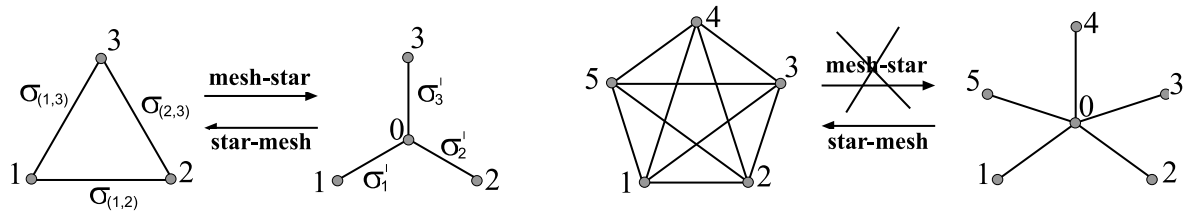}
\end{center}
\caption{A circuit built on a 3-simplex (a triangle) is always
equivalent to another circuit built on a star (left): this is the
star-triangle, or star-mesh, transformation. For a generic
$n$-simplex lattice (right) a generalized star-to-mesh
transformation is still valid, while the mesh-to-star transformation
holds only under Wheatstone's conditions.} \label{fig:star_mesh}
\end{figure}

Given the $1^{\mathrm{st}}$-generation fractal, the
$2^{\mathrm{nd}}$-generation fractal is built as follows. $n$
samples of generation 1 are arranged counterclockwise, and labelled
with the same numbers as the poles of the basic graph (see figure
\ref{fig:K5_generazione2}, left, for graph $K_5$). Then we connect
(short-circuit) pole $i$ of basic graph $j$ with pole $j$ of basic
graph $i$, with $j\neq i$; such connected poles are termed the
``internal poles'' of the resulting graph; the related potentials
are called $v_{(i,\,j)}$ (note that there is a one-to-one
correspondence between the internal potentials of the
$2^{\mathrm{nd}}$-order fractal and the links of the basic graph).
The poles that are left free from this procedure, i.e. every pole
$i$ of graph $i$, are the external poles of the
$2^{\mathrm{nd}}$-generation fractal, with their incoming currents
$I_i$ and external potentials $V_i$.\\
This circuit (figure \ref{fig:K5_generazione2}) is equivalent to
that built on a $1^{\,\mathrm{st}}$-generation fractal, whose
admittances we will call $\sigma_{(i,\,j)}^{(2)}$ (the equivalence
comes from the fact that both circuits have $n(n-1)/2$ degrees of
freedom). In general, $\sigma_{(i,\,j)}^{(g)}$ will be the
admittances of the $1^{\,\mathrm{st}}$-generation circuit equivalent
to that built on a $g^{\,\mathrm{th}}$-generation fractal with
$\sigma_{(i,\,j)}^{(1)}$ on the basic cell. We want to find the
decimation map $\mathcal{D}$ connecting the
$\{\sigma_{(i,\,j)}^{(1)}\}$ to the $\{\sigma_{(i,\,j)}^{(2)}\}$:
\begin{equation}
\vec\sigma^{(2)}=
\mathcal{D}\left(\vec\sigma^{(1)}\right),
\end{equation}
so that for a $g^{\,\mathrm{th}}$-generation fractal
$$\vec\sigma^{(g)}=
\mathcal{D}^{\,g}\left(\vec\sigma^{(1)}\right).$$
In 2 dimensions the decimation is carried out by means of the
star-delta (or star-mesh) transformations from electrostatics
\cite{Burioni2004} (figure \ref{fig:star_mesh}). Given a
triangle-shaped circuit, it is always possible to find an equivalent
star-shaped circuit, with an additional central pole, and new
admittances that depend on the old ones via easy algebraic
relations. The vice-versa (star-to-mesh) is also always possible.
For a general complete graph we can define a generalized version of
the star-mesh transformation: given a complete $K_n$ graph (a
generalized mesh), find the star graph with an additional hub pole
that is equivalent to the mesh
(mesh-to-star); or vice-versa (star-to-mesh).\\
It turns out, however \cite{Shen1947}, that for $n>3$ only the
star-to-mesh transformation is always possible, while the
mesh-to-star transformation is possible only under very restrictive
conditions. These are the so-called \textit{Wheatstone's
conditions}: in every 4-link subgraph of the mesh forming a
quadrilateral, the products of opposite admittances must be equal.
Since these conditions are not satisfied in general, a different
approach must be sought. Our approach, based on the direct
manipulation of the Laplacian matrix of the circuit, has been used
in \cite{Burioni2005} for $n=4$. Here we show a general method valid
for any $n$.
%
%
%
%
%
%
%
%
%
%
%
%
%
%
%
%
%
%
%
%
%
%
%
%
%
%
%
%

\section{Decimation: general method, fixed points and asymptotic expansion}
\label{Sec:Decimation}
\subsection{General procedure}
The Laplacian matrix of the $2^{\,\mathrm{nd}}$-generation fractal
is:
\begin{equation}
\mathbf{L^{(2)}}=\stackrel{
\overbrace{\qquad\qquad}^{\scriptstyle{i=1,\ldots,\,n}}
\overbrace{\qquad\qquad\qquad\quad\;\;}^{\scriptstyle{\alpha=1,\ldots,\,n(n-1)/2}}
}
{\left(\begin{array}{cccc|cccccccc}
& &  & & & & & & & \\
& & \mathbf{D}& &   & & & & \mathbf{\Sigma^T} & & & \\
& &  & & & & & & & \\
\hline  & &  & & & & & & & \\
  & & & &   \\
& & \mathbf{\Sigma}&  & & & & &\mathbf{M} & & &  \\
  & & & &   \\
  & & & &
\end{array}\right)
};
\end{equation}
We use an indexing such that the first $n$ indices of a row (or
column) refer to external points (and are labelled as the {\it
nodes} of the basic cell), the last $n(n-1)/2$ refer to internal
points (and are labelled as the {\it links}
of the basic cell).\\
$\mathbf{D}$ is a diagonal $n\times n$ matrix with entries
$D_{i\,i}=\sum_{j\neq i}\sigma_{(i,\,j)}^{(1)}$ (since an external
pole has no links with other external poles).\\
Matrix $\mathbf{\Sigma}$, that has internal poles labels (links of
the basic cell) as row indices and external poles labels (nodes of
the basic cell) as column indices, has only two entries different
from zero:
\begin{equation}\left\{
\begin{array}{ll}
\displaystyle{\Sigma_{\alpha\,k}=-\sigma_{\alpha}^{(1)}} &
\quad\mathrm{for\;\alpha\;incident\;in}\;k\mathrm{\;in\;the\;basic\;cell}\\[1ex]
\displaystyle{\Sigma_{\alpha\,k}=0} & \quad\mathrm{otherwise}\\[1ex]
\end{array}\right.
\end{equation}
$\mathbf{M}$ is a $[n(n-1)/2]\times [n(n-1)/2]$ symmetric matrix.
Its entries correspond to internal poles (hence,
to links of the basic cell). $\mathbf{M}$ has entries:\\
\begin{equation}\fl\left\{
\begin{array}{ll}
\displaystyle{M_{\alpha\,\alpha}=2\,\sigma_{\alpha}^{(1)}\,\,\,+\!\!\!\!
\sum_{\beta\mathrm{\;adjacent\;to \;}\alpha}\!\!\!\!\!\!\!\sigma_{\beta}^{(1)}} & \quad\mathrm{for\;diagonal\;entries}\\[2.5ex]
\displaystyle{M_{\alpha\,\beta}=-\sigma_{\beta}^{(1)}} & \quad\mathrm{if\;links\;\alpha\;and\;\beta\;of\;the\;basic\;cell\;are\;adjacent}\\[2.5ex]
\displaystyle{M_{\alpha\,\beta}=0} &
\quad\mathrm{if\;links\;\alpha\;and\;\beta\;of\;the\;basic\;cell\;are\;not\;adjacent.}\\[1ex]
\end{array}\right.
\end{equation}
The Kirchhoff's equations for the graph are:
\begin{displaymath}
\left(
\begin{array}{cc}
\mathbf{D} & \mathbf{\Sigma^T} \\
\mathbf{\Sigma} & \mathbf{M}
\end{array}
\right)
\left(\begin{array}{c}
\vec{V}\\\vec{v}
\end{array}\right)
=
\left(
\begin{array}{cc}
\mathbf{D}\vec{V} + \mathbf{\Sigma^T}\vec{v}\\
\mathbf{\Sigma}\vec{V}+\mathbf{M}\,\vec{v}
\end{array}
\right)
=
\left(\begin{array}{c}
\vec{I}\\0
\end{array}\right)
\end{displaymath}
Hence, we have two sets of equations. From the second set (regarding
the internal poles), we get
$\mathbf{M}\,\vec{v}=-\mathbf{\Sigma}\,\vec{V}$, and we can find the
internal potentials as functions of the $\sigma_{(i,\,j)}^{(1)}$:
$\vec{v}=-\mathbf{M^{-1}}\,\mathbf{\Sigma}\,\vec{V}$. Now we plug
this solution into the first set of equations and get
$(\mathbf{D}-\mathbf{\Sigma^T}\,\mathbf{M^{-1}}\,\mathbf{\Sigma})\vec{V}=\vec{I}$.
Hence, by comparison with equation (\ref{eq:laplacian_1stgen}), we
find that
$$\mathbf{L^{(1)}}=\mathbf{D}-\mathbf{\Sigma^T}\,\mathbf{M^{-1}}\,\mathbf{\Sigma},$$
where the left-hand side is understood to depend on the
$\sigma^{(2)}_{(i,\,j)}$, and the right-hand side on the
$\sigma^{(1)}_{(i,\,j)}$. The
decimation map $\mathcal{D}$ is given by the off-diagonal entries:
\begin{equation}
\sigma^{(2)}_{(i,\,j)}=-(\mathbf{L^{(1)}})_{i\,j}=(\mathbf{\Sigma^T}\,\mathbf{M^{-1}}\,\mathbf{\Sigma})_{i\,j}.
\label{eq:decimation_sigma2}
\end{equation}
In general, $(M^{-1})_{\alpha\,\beta}=P_{\alpha\,\beta}/\Delta$,
where $P_{\alpha\beta}$ is a homogeneous polynomial of
degree $n(n-3)/2$ in the variables $\{\sigma_{(i,\,j)}\}$, while
$\Delta=\mathrm{det}(\mathbf{M})$ is a homogenous polynomial of
degree $n(n-1)/2$ in the variables $\{\sigma_{(i,\,j)}\}$. \\
From these considerations it follows that $\mathcal{D}$ is a
$n(n-1)/2$-dimensional rational map from
$\mathbb{C}_{\,\infty}^{n(n-1)/2}$ to
$\mathbb{C}_{\,\infty}^{n(n-1)/2}$ in the variables
$\{\sigma_{(i,\,j)}\}$, where
$\mathbb{C}_{\,\infty}=\mathbb{C}\cup\{\infty\}$; we refer the
reader to the specific literature \cite{Blanchard1984} for the
language of dynamical systems and rational maps. By construction,
$\mathcal{D}$ is homogeneous of degree 1 in its variables:
$\mathcal{D}(\lambda\,
\vec{\sigma})=\lambda\,\mathcal{D}(\vec{\sigma})$, $\lambda \in
\mathbb{C}$. This property allows us to recover a physical meaning
for those points with a negative real value (the physical constraint
on the admittance $\sigma$ of a passive element being that
$\mbox{Re}(\sigma)\geq0$). Indeed, if a result holds for a
$\vec{\sigma}$, it also holds for all the $\lambda\,\vec{\sigma}$,
$\lambda \in \mathbb{C}$. Hence, a vector $\vec{\sigma}$ is
``physically meaningful'' provided that there exists a $\lambda \in
\mathbb{C}$ such that $\mbox{Re}(\lambda
\sigma_{\alpha})\geq0\,\forall\alpha$. For instance, a vector such
as $\vec{\sigma}=(-1,1,1,\ldots,\,1)$ makes sense since it can be
multiplied, e.g., for $i$ to give $(-i,i,i,\ldots,\,i)$ (a set of
capacitive and inductive admittances), while the vector
$(1+i,1-i,-1,\ldots)$ cannot be mapped by multiplication into any
physically meaningful point. The requirement a set of admittances
must satisfy to have a physical meaning is that the set of the
vectors representing them in the complex plane cover an angle $\leq 180^{\circ}$.\\
The number of terms in expression (\ref{eq:decimation_sigma2}) grows
very fast with $n$ (approximately as $n!$): for example, the number
of terms in $\Delta=\mathrm{det}(\mathbf{M})$ is 7 for $n=3$, 293
for $n=4$, 61763 for $n=5$. Hence, the map is impossible to study as
it is. We will focus here on its asymptotic properties. First, we
can show that the \textit{isotropic point at 0} is an attracting
fixed point for $\mathcal{D}$. Second, by perturbation of the
isotropic solution, we can find the exact asymptotic expansion of
$\mathcal{D}$ near the fixed point for any $n$.
%
%
%
%
%
%
%
%
%
%
%
%
%
%
%
%
%
%
%
%
%
%
%
%
\subsection{Isotropic fractal and fixed points}
\label{Sec:isotropic_fractal}
In the isotropic fractal all the admittances have the same value:
\[\sigma_{(i,\,j)}=\sigma_0\qquad \forall i,j.
\]
Matrix $\mathbf{M}\equiv \mathbf{M_0}$ gets the general form:\\
\begin{displaymath}\left\{
\begin{array}{ll}
\displaystyle{({M_0})_{\alpha\,\alpha}=2(n-1)\sigma_0} & \qquad\mathrm{for\;diagonal\;entries}\\[1.2ex]
\displaystyle{({M_0})_{\alpha\,\beta}=-\sigma_0} & \qquad\mathrm{\alpha\;and\;\beta\;adjacent}\\[1.2ex]
\displaystyle{({M_0})_{\alpha\,\beta}=0} &
\qquad\mathrm{\alpha\;and\;\beta\;not\;adjacent.}\\[.5ex]
\end{array}\right.
\end{displaymath}
The general form of its inverse $\mathbf{M_0^{-1}}$ is \\
\begin{displaymath}\left\{
\begin{array}{ll}
\displaystyle{\left({M_0}^{-1}\right)_{\alpha\,\alpha}=\frac{n+6}{2n(n+2)}\sigma_0^{\,-1}} & \qquad\mathrm{diagonal\;entries}\\[2ex]
\displaystyle{\left({M_0}^{-1}\right)_{\alpha\,\beta}=\frac{3}{2n(n+2)}\sigma_0^{\,-1}} & \qquad\mathrm{\alpha\;and\;\beta\;adjacent}\\[1.5ex]
\displaystyle{\left({M_0}^{-1}\right)_{\alpha\,\beta}=\frac{2}{2n(n+2)}\sigma_0^{\,-1}} &
\qquad\mathrm{\alpha\;and\;\beta\;not\;adjacent,}\\[2ex]
\end{array}\right.
\end{displaymath}
as can be checked by direct multiplication.
The general form of $-\mathbf{L_0^{(1)}}=\mathbf{\Sigma_0^T}\,\mathbf{M_0^{-1}}\,\mathbf{\Sigma_0}$ is \\
\begin{displaymath}\left\{
\begin{array}{ll}
\displaystyle{(-\mathbf{L_0^{(1)}})_{i\,i}=\frac{2n-2}{n+2}\,\sigma_0} & \qquad\mathrm{diagonal\;entries}\\[2ex]
\displaystyle{(-\mathbf{L_0^{(1)}})_{i\,j}=\frac{n}{n+2}\,\sigma_0} & \qquad\mathrm{off-diagonal\;entries.}\\[2ex]
\end{array}\right.
\end{displaymath}
By looking at the off-diagonal entries of $-\mathbf{L_0^{(1)}}$, we
find that $\mathcal{D}$ maps isotropic vectors into isotropic
vectors with a scaling factor of $n/(n+2)$:
\begin{displaymath}
\mathcal{D}(\vec{\sigma}_0)= \frac{n}{n+2}\vec{\sigma}_0,
\label{eq:isotropic_law}
\end{displaymath}
where $\vec{\sigma}_0=(\sigma_0,\,\sigma_0, \ldots, \,\sigma_0)$.
The isotropic point at 0 is an attractor for map $\mathcal{D}$,
with the leading term of the asymptotic behaviour:
\begin{equation}
\sigma_{\alpha}^{(g)}\sim\left(\frac{n}{n+2}\right)^g \sigma_0\qquad
\forall\alpha\;\;\mathrm{for}\;g\rightarrow\infty,
\end{equation}
where $g$ is the generation of the fractal and $\sigma_0$ depends on
the initial conditions. Recalling that the linear size of the
fractal scales as $2^g$, the exponent $\bar{\zeta}=\ln[(n+2)/n]/\ln
2$ for the scaling of the isotropic conductance introduced in
section \ref{Sec:Introduction}
is recovered.\\
%
%
%
%
%
%
%
%
%
%
%
%
%
%
%
%
%
%
%
%
%
%
%
%
\subsection{Perturbation of the isotropic fractal and asymptotic expansion}
\label{Sec:perturbation}
We now add a small perturbation to the isotropic fractal by
incrementing variable $\sigma_{(1,\,2)}$ by a value $\varepsilon$:
\begin{displaymath}
\sigma_{(1,\,2)}=\sigma_0+\varepsilon,\quad\mathrm{while}\,\sigma_{(i,\,j)}=\sigma_0\quad\mathrm{for}(i,\,j)\neq(1,\,2),
\end{displaymath}
with $\varepsilon/\sigma_0$ small. $\mathbf{M}$ and
$\mathbf{\Sigma}$ are increased by infinitesimal matrices:
$\mathbf{M}\,=\,\mathbf{M_{0}}\,+\,\mathbf{\delta M}$;
$\mathbf{\Sigma}\,=\,\mathbf{\Sigma_{0}}\,+\,\mathbf{\delta
\Sigma}$, and the equation for $-\mathbf{L^{(1)}}$ becomes
\begin{eqnarray}
-\mathbf{L^{(1)}}=\mathbf{\Sigma^T}\,\mathbf{M^{-1}}\mathbf{\Sigma}\sim
\left(\mathbf{\Sigma_0}+\mathbf{\delta\Sigma}\right)^T
\left(\mathbf{M_0}+\mathbf{\delta M}\right)^{-1}
\left(\mathbf{\Sigma_0}+\mathbf{\delta\Sigma}\right)
\\[1ex]\nonumber
\sim\left(\mathbf{\Sigma_0^T}+\mathbf{\delta\Sigma^T}\right)
\left(\mathbf{I}-\mathbf{M_0^{-1}}\mathbf{\delta M}\right)\mathbf{M_0^{-1}}
\left(\mathbf{\Sigma_0}+\mathbf{\delta\Sigma}\right)
\\[1ex]\nonumber
\sim\underbrace{\mathbf{\Sigma_0^T}\,\mathbf{M_0^{-1}}\mathbf{\Sigma_0}}_{\mathrm{isotropic\;solution}}+
\underbrace{\mathbf{\Sigma_0^T}\,\mathbf{M_0^{-1}}\mathbf{\delta\Sigma}
+\mathbf{\delta\Sigma^T}\,\mathbf{M_0^{-1}}\mathbf{\Sigma_0}
-\mathbf{\Sigma_0^T}\,\mathbf{M_0^{-1}}\,\mathbf{\delta M}\,\mathbf{M_0^{-1}}\mathbf{\Sigma_0}}_{\mathrm{perturbation}}
\\[1ex]\nonumber
=-\mathbf{L_0^{(1)}}-\mathbf{\delta L^{(1)}}.
\label{eq:delta_v_general}
\end{eqnarray}
Knowing the explicit expressions of $\mathbf{\delta M}$ and
$\mathbf{\delta \Sigma}$, the calculation of the terms contributing
to $-\mathbf{\delta L^{(1)}}$ is lengthy but not difficult. We omit
here the details and give only the final result. $-\mathbf{\delta
L^{(1)}}$ is a symmetric $n\times n$ matrix with entries:
\begin{displaymath}\left\{
\begin{array}{ll}
\displaystyle{-\frac{3n+2}{(n+2)^2}}\,\varepsilon & \qquad\mathrm{diagonal}\;\mathrm{entries}\;(1,1)\;\mathrm{and}\;(2,2)\\[3ex]
\displaystyle{\frac{5n-2}{(n+2)^2}}\,\varepsilon & \qquad\mathrm{entries}\;(1,2)\;\mathrm{and}\;(2,1)\\[3ex]
\displaystyle{\frac{n-2}{(n+2)^2}}\,\varepsilon & \qquad\mathrm{entries}\;(1,i)\;\mathrm{and}\;(2,i),\;i\neq 1,2\\[3ex]
\displaystyle{\frac{-2}{(n+2)^2}}\,\varepsilon & \qquad\mathrm{entries}\;(i,j),\;\mathrm{with}\;i,j\neq 1,2.\\[2ex]
\end{array}\right.
\end{displaymath}
By looking at the off-diagonal entries, we see that by incrementing
$\sigma_{(1,\,2)}$ the admittances undergo the following changes:
\begin{displaymath}\left\{
\begin{array}{l}
\displaystyle{\sigma_{(1,\,2)}:\qquad\qquad\sigma_0+\varepsilon
\longrightarrow\frac{n}{n+2}\sigma_0+\frac{5n-2}{(n+2)^2}\,\varepsilon}\\[2,5ex]
\displaystyle{\sigma_{(1,\,\,i)},\,\sigma_{(2,\,\,i)}:\,\,\quad\sigma_0\:\:
\longrightarrow\:\: \frac{n}{n+2}\sigma_0+\frac{n-2}{(n+2)^2}\,\varepsilon}\\[2,5ex]
\displaystyle{\sigma_{(i,\,j)}:\qquad\qquad\,\sigma_0\:\:
\longrightarrow\:\: \frac{n}{n+2}\sigma_0-\frac{2}{(n+2)^2}\,\varepsilon}.\\[2,5ex]
\end{array}\right.
\end{displaymath}
This means that in general, by incrementing \textit{each} admittance
$\sigma_{\alpha}$ by its own infinitesimal quantity near the
isotropic point
$\sigma_{\alpha}\rightarrow\sigma_0+\varepsilon_{\alpha}$ (in vector
notation $\vec{\sigma}\rightarrow\vec{\sigma}_0+\vec{\varepsilon}$),
and labelling with $\alpha'$ the links adjacent to $\alpha$, with
$\alpha''$ the links not
adjacent to $\alpha$, the map changes the increments in the following way:\\
\begin{equation*}
\varepsilon_{\alpha}\;\longrightarrow\;\underbrace{\frac{5n-2}{(n+2)^2}\,
\varepsilon_{\alpha}}_{\mathrm{diagonal\;term}}\;+\;\sum_{\alpha'}
\underbrace{\frac{n-2}{(n+2)^2}\,\varepsilon_{\alpha'}}_{\mathrm{adjacent\;links }}\;
+\;\sum_{\alpha''}\underbrace{\frac{-2}{(n+2)^2}\,\varepsilon_{\alpha''}}_{\mathrm{ non-adjacent\;links}}.
\end{equation*}
In conclusion, in the vicinity of the isotropic point $\mathcal{D}$
gets the linearized form:
\begin{equation}
\mathcal{D}\left(\vec{\sigma}_0+\vec{\varepsilon}\right)\sim\frac{n}{n+2}\,\vec{\sigma}_0+\mathbf{E}\,\vec{\varepsilon},
\end{equation}
where $\mathbf{E}$ is a perturbation matrix with entries:\\
\begin{equation}\left\{
\begin{array}{ll}
\displaystyle{E_{\,\alpha\,\alpha}=\frac{5n-2}{(n+2)^2}} & \qquad\mathrm{diagonal\;entries}\\[2ex]
\displaystyle{E_{\,\alpha\,\beta}=\frac{n-2}{(n+2)^2}} &  \qquad\alpha\;\mathrm{and}\;\beta\;\mathrm{adjacent}\\[2ex]
\displaystyle{E_{\,\alpha\,\beta}=}\frac{-2}{(n+2)^2} & \qquad\alpha\;\mathrm{and}\;\beta\;\mathrm{not\;adjacent.}\\[2ex]
\end{array}\right.
\end{equation}
The following section is devoted to the study of the eigenvalues and
eigenspaces of $\mathbf{E}$.
%
%
%
%
%
%
%
%
%
%
%
%
%
%
%
%
%
%
%
%
%
%
%
%
\section{Eigenvalues and eigenspaces of the linearized map}
\label{Sec:Eigenvalues}
The basic result is that matrix $\mathbf{E}$ has three eigenvalues
$e_1$, $e_2$, $e_3$ for any $n$ (with $e_3$ disappearing only for
$n=3$). Their values and multiplicities are:
\begin{equation*}
\begin{array}{ll}
\mathrm{eigenvalue} & \qquad\mathrm{multiplicity}\\[1,5ex]
\displaystyle{e_1=\frac{n}{n+2}} & \qquad\displaystyle{1}\\[3ex]
\displaystyle{e_2=\frac{n\,(n+1)}{(n+2)^2}} & \qquad\displaystyle{n-1}\\[3ex]
\displaystyle{e_3=\frac{3\,n}{(n+2)^2}} & \qquad\displaystyle{\frac{1}{2\,}n\,(n-3)}\\[-1ex]
\label{eq:eigenvalues_sigma}
\end{array}
\end{equation*}
We call $\mathcal{E}_1$, $\mathcal{E}_2$, $\mathcal{E}_3$ the
subspaces spanned by $e_1$, $e_2$,
$e_3$, respectively.\\
The isotropic vector with $1$ on all the links is an eigenvector of
$\mathbf{E}$ (since all rows are permutations of each other). The
eigenvalue is the sum of the entries of a row, that is easily shown
to be $\frac{n}{n+2}$. This could be expected from equation
(\ref{eq:isotropic_law}), since an equal increment of $\varepsilon$
for all admittances just shifts the system to
the different isotropic point $\sigma_0+\varepsilon$. \\
\begin{figure}
\begin{center}
\includegraphics[width=0.8\textwidth]{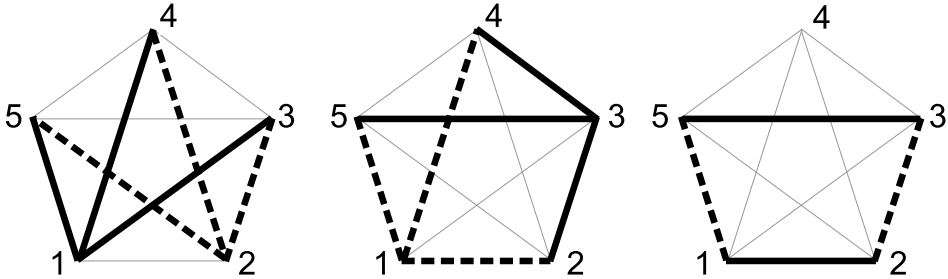}
\end{center}
\caption{Some eigenvectors of the matrix $\mathbf{E}$ of the
linearized map for the $K_5$ graph: thin lines denote links with
value 0, thick continuous lines denote links with value 1, and thick
dotted lines links with value -1. The vectors
$\vec{\mathsf{f}}_{\,(1,\,2)}$ and $\vec{\mathsf{f}}_{\,(1,\,3)}$,
that are eigenvectors for the eigenvalue $e_2$, are shown on the
first two graphs from the left. A 4-loop with alternating 1 and -1
on its links is shown on the graph on the right: it is an
eigenvector for eigenvalue $e_3$.}
\label{fig:eigenvect_subleading_K5}
\end{figure}
\subsection{Eigenvectors corresponding to $e_2$: Wheatstone's conditions}
\begin{figure}
\begin{center}
\includegraphics[width=0.8\textwidth]{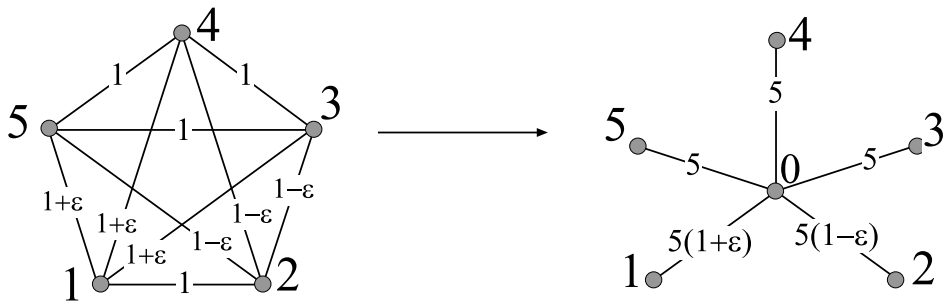}
\end{center}
\caption{A $K_5$ graph with $\sigma_0=1$ perturbed with an
eigenvector $\varepsilon\vec{\mathsf{f}}_{\,(1,\,2)}$ corresponding
to eigenvalue $e_2$ ({\it left}). The circuit satisfies Wheatstone's
conditions to first order in $\varepsilon$: in every subgraph
forming a quadrilateral the products of opposite admittances are
equal. A generalized mesh-star transformation is thus possible ({\it
right}).} \label{fig:star_mesh_K5}
\end{figure}

The eigenvectors for eigenvalue $e_2$ can be built as follows. Take
two vertices of the graph, for instance 1 and 2, and consider the
link $(1,\,2)$ joining them. Then build a vector this way: all the
$n-2$ links incident with 1, \textit{except} link $(1,\,2)$, are
given the value $1$; all the $n-2$ links incident with node 2,
except link $(1,\,2)$, are given the value $-1$; link $(1,\,2)$ and
all the other links have value $0$. We call
$\vec{\mathsf{f}}_{\,(1,\,2)}$ this vector; in general we call
$\vec{\mathsf{f}}_{\,(i,\,j)}$ the vector built in this way starting
from link $(i,\,j)$. Figure \ref{fig:eigenvect_subleading_K5} shows
two of these eigenvectors
for the graph $K_5$.\\
$\vec{\mathsf{f}}_{\,(i,\,j)}$ can be shown to be an eigenvector for
matrix $\mathbf{E}$ by direct multiplication for the matrix. An
easier and physically more interesting way is seeing that for
vectors $\vec\sigma_0+\varepsilon\,\vec{\mathsf{f}}_{\,(i,\,j)}$ the
Wheatstone's conditions mentioned in section
\ref{Sec:the_basic_cell} hold to first order, hence, a generalized
mesh-star transformation is valid (figure \ref{fig:star_mesh_K5}).
The star equivalent to this mesh has an admittance equal to
$n\,(\sigma_0+\varepsilon)$ for links $(0,\,i)$ (0 being the
additional central node), $n\,(\sigma_0-\varepsilon)$ for links
$(0,\,j)$, and $n\,\sigma_0$ on all the other links. The decimation
can then be carried out with the same easy topological procedures
described in \cite{Burioni2004} for the 3-simplex lattice, yielding
$\frac{n}{n+2}\vec\sigma_0+\frac{n(n+1)}{(n+2)^2}\,\varepsilon\,\vec{\mathsf{f}}_{\,(i,\,j)}$.\\
It can be easily shown that the set $\vec{\mathsf{f}}_{\,(1,\,j)}$
(with $i=1$ fixed and $j=2,\ldots,\,n$) is a set of linearly
independent (not orthogonal) vectors. Furthermore, any other
$\vec{\mathsf{f}}_{\,(k,\,l)}$ can be obtained from this set by
means of the formula
$\vec{\mathsf{f}}_{\,(k,\,l)}=\vec{\mathsf{f}}_{\,(1,\,l)}-\vec{\mathsf{f}}_{\,(1,\,k)}$.
Hence, the number of independent vectors of this eigenspace
(the multiplicity of the eigenvalue $e_2$) is $n-1$.\\
The eigenvalue $e_2$ corresponds to the well-known anisotropy
exponent $\bar{\lambda}=\ln(e_1/e_2)/\ln 2=\ln[(n+2)/(n+1)]/\ln 2$
discussed in the Introduction. However, the dimension of the
eigenspace $\mathcal{E}_2$ and the Wheatstone property where not
known up to now.
\subsection{Eigenvectors corresponding to $e_3$: even-cycle space and crossover property}
\begin{figure}[b]
\begin{center}
\includegraphics[width=0.7\textwidth]{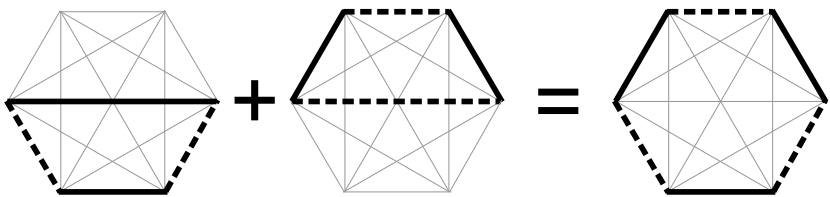}
\end{center}
\caption{\small A 6-cycle be obtained from two 4-cycles in the way
shown above in a $K_6$ graph. The same is true for any even cycle:
4-cycles form a good basis for the space $\mathcal{E}_3$.
\label{fig:4+4=6}}
\end{figure}

In graph theory, a \textit{cycle} of length $\ell$, or $\ell$-cycle,
is a closed walk (sequence of connected vertices) composed of $\ell$
distinct points and links. The eigenvectors for eigenvalue $e_3$ are
\textit{all the cycles of even length with alternating 1 and -1 on
their links}, and 0 on the links not belonging to the cycle (we will
call them simply ``even cycles'' henceforth, with the implicit
assumption on the value of their links). One such eigenvector for
$K_5$ is shown in figure \ref{fig:eigenvect_subleading_K5}, on the
right, in the case of a 4-cycle. In this case, the property of being
an eigenvector for $\mathbf{E}$ has to be
verified by direct multiplication.\\
The eigenvalue $e_3$ corresponds to a {\it secondary} anisotropy
exponent with value $\bar\mu=\ln(e_1/e_3)/\ln 2=\ln[(n+2)/3]/\ln 2$.
The new exponent appears when the contribution of the first
anisotropy exponent $\bar\lambda$ is negligible, that is, if we
start from a configuration
orthogonal (or almost orthogonal) to the space $\mathcal{E}_2$.\\
Since a generic even cycle can be obtained from a linear combination
of 4-cycles (figure \ref{fig:4+4=6}), 4-cycles are a basis for the
space $\mathcal{E}_3$. The number $\mathcal{N}_{4,\,n}$ of linearly
independent 4-cycles on a $K_n$ graph (the multiplicity of $e_3$)
can be computed (for example, by recurrence starting from a $K_4$
graph) and is found to be $\mathcal{N}_{4,\,n}=\frac{1}{2}n(n-3)$.
For $n=3$, when the basic cell is a triangle (and the fractal is the
usual two-dimensional Sierpi\'{n}ski gasket), there are no even
loops and the only eigenvalues are $e_1$ and $e_2$
\cite{Burioni2004}. For $n\geq5$, the eigenspace $\mathcal{E}_3$ is
that with the largest dimensionality,
growing as $n^2$ for $n\gg 1$, as opposed to that of $\mathcal{E}_2$ that grows as $n$.\\
When the starting configuration is not exactly orthogonal to
$\mathcal{E}_2$, the behaviour of the projection of $\vec\sigma$ onto
$\mathcal{E}_3$ displays an interesting crossover property, due to
the homogeneity of the map $\mathcal{D}$, that makes the value of
$\bar\mu$ deviate from that calculated in the first-order
approximation. In order to see it, we need to calculate the
second-order term in the series expansion for the projection of
$\vec\sigma$ onto $\mathcal{E}_3$: it typically contains quadratic
contributions of the form
$$(\ldots)\frac{\sigma_{(1,\,2)}^2}{\sigma_0}+(\ldots)\frac{\sigma_{(1,\,2)}\,\sigma_{(1,\,3)}}{\sigma_0}+...,$$
where the $(\ldots)$ denote some constant. Due to the $\sigma_0$ in
the denominator, for $g\rightarrow\infty$, the leading term of this
expansion is proportional to
$$\left(\frac{e_2^2}{e_1}\right)^g=\left(\frac{n(n+1)^2}{(n+2)^3}\right)^g>(e_3)^g.$$
Hence, the asymptotic behaviour $\sim(e_3)^g$ normally holds only for
some iterations (the number of which depend on the initial
conditions), then the second-order term $\sim(e_2^2/e_1)^g$
prevails. This means that the second anisotropy exponent $\bar\mu$
crosses over from $\bar\mu=\ln[(n+2)/3]/\ln 2$ to
$\bar\mu=\ln[(n+2)^3/n(n+1)^2]/\ln 2$.
%
%
%
%
%
%
%
%
%
%
%
%
%
%
%
%
%
%
%
%
%
%
%
%
%
%
\section{Conclusions}
\label{Sec:Conclusions}
We have extended the method used in \cite{Burioni2005} to find the
decimation map $\mathcal{D}$ for a network of admittances on a
generic $n$-simplex lattice. By means of a linear expansion near the
isotropic fixed point of the map, we have found the first-order
asymptotic behaviour of $\mathcal{D}$ for every $n$. The eigenspaces
of the linearized map have been found to be always three (two just
for $n=3$), and to have a direct interpretation in terms of graph
theory. In particular, the third eigenspace is connected to the set
of even-length cycles on the basic cell, and its eigenvalue is
related to a secondary anisotropy exponent $\bar\mu$ with a value
that crosses over from $\ln[(n+2)/3]/\ln 2$ to
$\ln[(n+2)^3/n(n+1)^2]/\ln 2$
with the size of the fractal.\\
Due to a well-established correspondence between electrical networks
and random walks \cite{Doyle1984}, our results could be easily
extended to the random walk problem on these lattices, with jumping
probabilities depending on the direction. Since our method of direct
manipulation of the Laplacian matrix is quite general, we are
currently studying its application to other exactly-decimable
fractals, to non-exactly-decimable fractals (with some proper
approximations), and to more general graphs.

\end{document}